\documentclass[10pt,sigconf]{acmart}
\AtBeginDocument{%
  \providecommand\BibTeX{{%
    \normalfont B\kern-0.5em{\scshape i\kern-0.25em b}\kern-0.8em\TeX}}}

\usepackage{listings}
\usepackage{amsmath}
\usepackage{multirow}

\lstdefinelanguage{Julia}%
  {morekeywords={abstract,break,case,catch,const,continue,do,else,elseif,end,export,false,for,function,immutable,import,importall,if,in,macro,module,otherwise,quote,return,switch,true,try,type,typealias,using,while, begin, @inbounds, nothing},%
   sensitive=true,%
   morecomment=[l]\#,%
   morecomment=[n]{\#=}{=\#},%
   morestring=[s]{"}{"},%
   morestring=[m]{'}{'},%
}[keywords,comments,strings]%

\copyrightyear{2023} 
\acmYear{2023} 
\setcopyright{licensedusgovmixed}\acmConference[SC-W 2023]{Workshops of The International Conference on High Performance Computing, Network, Storage, and Analysis}{November 12--17, 2023}{Denver, CO, USA}
\acmBooktitle{Workshops of The International Conference on High Performance Computing, Network, Storage, and Analysis (SC-W 2023), November 12--17, 2023, Denver, CO, USA}
\acmPrice{15.00}
\acmDOI{10.1145/3624062.3624278}
\acmISBN{979-8-4007-0785-8/23/11}

\begin{document}

\title[Julia as a Unifying End-to-End Workflow Language on the Frontier Exascale System]{Julia as a Unifying End-to-End Workflow Language \\ on the Frontier Exascale System}

\author{William F. Godoy, Pedro Valero-Lara, Caira Anderson, Katrina W. Lee, Ana Gainaru, Rafael Ferreira da Silva, and Jeffrey S. Vetter} 
\affiliation{ \institution{Oak Ridge National Laboratory\thanks{This manuscript has been authored by UT-Battelle LLC under contract DE-AC05-00OR22725 with the US Department of Energy (DOE). The publisher acknowledges the US government license to provide public access under the DOE Public Access Plan (\url{https://energy.gov/downloads/doe-public-access-plan}).}} 
\city{Oak Ridge}
\state{TN}
\country{USA} }
\email{{godoywf,valerolarap,andersonci,leekw,gainarua,silvarf,vetter}@ornl.gov}

\renewcommand{\shortauthors}{Godoy, et al.}

\begin{abstract}
We evaluate Julia as a single language and ecosystem paradigm powered by LLVM to develop workflow components for high-performance computing. We run a Gray-Scott, 2-variable diffusion-reaction application using a memory-bound, 7-point stencil kernel on Frontier, the US Department of Energy’s first exascale supercomputer. We evaluate the performance, scaling, and trade-offs of (i) the computational kernel on AMD’s MI250x GPUs, (ii) weak scaling up to 4,096 MPI processes/GPUs or 512 nodes, (iii) parallel I/O writes using the ADIOS2 library bindings, and (iv) Jupyter Notebooks for interactive analysis. Results suggest that although Julia generates a reasonable LLVM-IR, a nearly 50\% performance difference exists vs. native AMD HIP stencil codes when running on the GPUs. As expected, we observed near-zero overhead when using MPI and parallel I/O bindings for system-wide installed implementations. Consequently, Julia emerges as a compelling high-performance and high-productivity workflow composition language, as measured on the fastest supercomputer in the world.
\end{abstract}




\maketitle

\section{Introduction}

% Discuss workflows and applications
Recent emphasis on the end-to-end workflow development process for high-performance computing (HPC) applications acknowledges the increasing complexity required for achieving performance, portability, and productivity~\cite{9309042}. This complexity is primarily driven by two factors: (i)~the evolving application requirements for experimental, observational, and computational science and (ii)~the extreme heterogeneity of our computing and data generation and processing systems~\cite{ferreiradasilva-fgcs-2017, doi:10.1177/1094342017704893,osti_1473756}. Consequently, the community vision for a research and development roadmap~\cite{9652570, wcs2022} has identified key challenges posed by integrating HPC, AI, and FAIR (findable, accessible, interoperable, and reusable)~\cite{10023945, goble2020fair} workflows at exascale.

% Talk about Julia
The Julia programming language~\cite{Bezanson2017-ca} was designed to provide a powerful unifying strategy to close the gaps between scientific computing and data science. 
Julia unifies aspects of the scientific workflow development process: simulation, communication, visualization, parallel data I/O, AI, and interactive computing. 
This is achieved by providing (i)~a dynamic just-in-time (JIT) compiled front end to LLVM~\cite{lattner2004llvm},  (ii)~a lightweight interoperability layer for existing C and Fortran HPC codes, and (iii)~a unified community ecosystem (e.g., packaging and testing). The status quo in HPC software is to write simulation code by using a compiled language (Fortran, C, or C\texttt{++}) and the Message Passing Interface (MPI)~\cite{gropp1999using} alongside on-node standard (OpenMP~\cite{openmp}), vendor (CUDA~\cite{cuda}, HIP~\cite{hip}) or third-party (Kokkos~\cite{Kokkos}, RAJA~\cite{Raja}) parallel programming models, also known as \texttt{MPI+X}, while data analysis and workflow orchestration tasks are programmed by using high-level interfaces and languages (e.g., Python)---not without caveats and trade-offs~\cite{9307940}. As a result, Julia provides a valuable alternative in the convergence of high-productivity and high-performance that must be tested on exascale hardware.

In this work, we measure and analyze the computational performance aspects of a Gray-Scott diffusion-reaction HPC workflow application~\cite{doi:10.1126/science.261.5118.189} written in Julia and running on Frontier, the recently deployed exascale system at Oak Ridge National Laboratory.\footnote{\url{https://www.olcf.ornl.gov/frontier}} We focus on the computational trade-offs of solving a representative application of a 7-point stencil kernel on Frontier's AMD's MI250x GPUs at different scales. Simultaneously, we focus on the production and consumption of data in Jupyter Notebooks through the parallel file system.
More specifically, we present the following:

\begin{itemize}
    \item Compute measurements on AMD MI250x GPUs and comparison with a similar HIP 7-point stencil kernel
    \item Weak scaling on up to 4,092 GPU (or 512 nodes) using MPI showing the variability on local process time-to-solution
    \item Parallel I/O scaling results using the ADIOS-2 Julia bindings
    \item The overall feasibility of Julia as a single language and ecosystem for all above components and its use on JupyterHub for data analysis of the generated data.
\end{itemize}

% Paper
The paper is organized as follows: Section~\ref{sec:Related Work} presents related work in the use of Julia in HPC environments. Section~\ref{sec:Workflow} describes the components of the tested workflow, including parallel simulations and data analysis. Section~\ref{sec:Implementation} describes the Julia implementation as a single unifying paradigm for GPU, MPI, parallel I/O, and data analysis using Frontier's compute nodes and the Lustre file system with the ADIOS2~\cite{GODOY2020100561} library. 
Section~\ref{sec:Results} shows performance and weak scaling results when using up to 4,092 GPUs or 512 nodes of the Frontier supercomputer, including achieved GPU bandwidths, weak scaling, and write performance for large simulations. Conclusions and future directions are presented in Section~\ref{sec:Conclusions}. Appendix~\ref{ap1:Artifact} presents the Artifact Description for the reproducibility of this study.

\section{Related Work}
\label{sec:Related Work}

Recent work has reported the use of Julia in different aspects of HPC, and its use is still an area of active exploration and community engagement~\cite{JuliaHPC}.

% Compute
On the compute side, Lin and McIntosh-Smith \citeN{9652798} compared memory-bound (BabelStream) and \\ compute-bound (miniBUDE) benchmark mini-app ports in Julia against several implementations on different CPU and GPU vendor systems. They concluded that Julia's on-node performance is either on par or slightly behind the vendor compiled implementations for their use cases. Godoy et al.~\citeN{10196600} showed that Julia is competitive for a simple matrix multiplication kernel on modern exascale nodes when compared alongside the high-level Python's Numba~\cite{lam2015numba} JIT layer running on top of LLVM and the C\texttt{++} Kokkos programming model against vendor OpenMP on CPU and vendor CUDA and HIP implementations on GPU. Faingnaert et al.~\citeN{9655458} provide optimized GEMM kernels in Julia that are competitive with cuBLAS and CUTLASS implementations. Ranocha et al. ~\citeN{Ranocha2022} describe similar levels of performance for the  \texttt{Trixi.jl} partial differential equation~(PDE) solver at scale when the Julia implementation is compared against compiled HPC languages.

% Network
On the network communication side, Giordano et al.~\citeN{9912702} measured comparable system-level scalability performance for Julia's MPI.jl~\cite{Byrne2021} bindings on the Arm-based Fujitsu A64FX Fugaku system running point-to-point communication benchmarks for different floating point representations. Hunold et al.\citeN{Hunold2020} used a Julia port of the STREAM benchmark~\cite{mccalpin1995memory} to show negligible overheads in the collective \texttt{Bcast}, \texttt{Alltoall}, and \texttt{Allreduce} MPI operations when compared against equivalent C implementations on up to 1,152 processes.
Shang et al. ~\citeN{10046049} used the Julia language ecosystem on the many-cores Sunway supercomputer to conduct quantum computational chemistry simulations. In their experiments they achieved up to 91\% efficiency when measuring weak scaling on up to 21M cores. Regier et al. ~\citeN{REGIER201989} provided Celeste, an early application of Julia on multi-threaded CPU HPC systems. Celeste achieved 1.54 petaflops using 1.3 million threads on 9,300 Knights Landing (KNL) nodes of the Cori supercomputer, while exploiting burst buffer for data I/O.
More recently, Lu et al.~\citeN{9912710} explored the integration of Julia in their framework for moving computation as data in binary or LLVM bit code representations. They found that their implementation can leverage JIT optimizations, but further exploration is needed.

To the best of our knowledge, our work represents the pioneering effort to harness the potential of Julia in modern HPC end-to-end workflows targeting a GPU-powered exascale system.
\vspace{-0.5cm}
\section{Workflow Description}
\label{sec:Workflow}

This section describes the characteristics of the computation, communication, parallel data I/O, and data analysis of the representative Gray-Scott diffusion-reaction simulation code running on Frontier.

\begin{figure}[!ht]
  \centering
  \includegraphics[width=1\columnwidth,height=1.1\columnwidth]{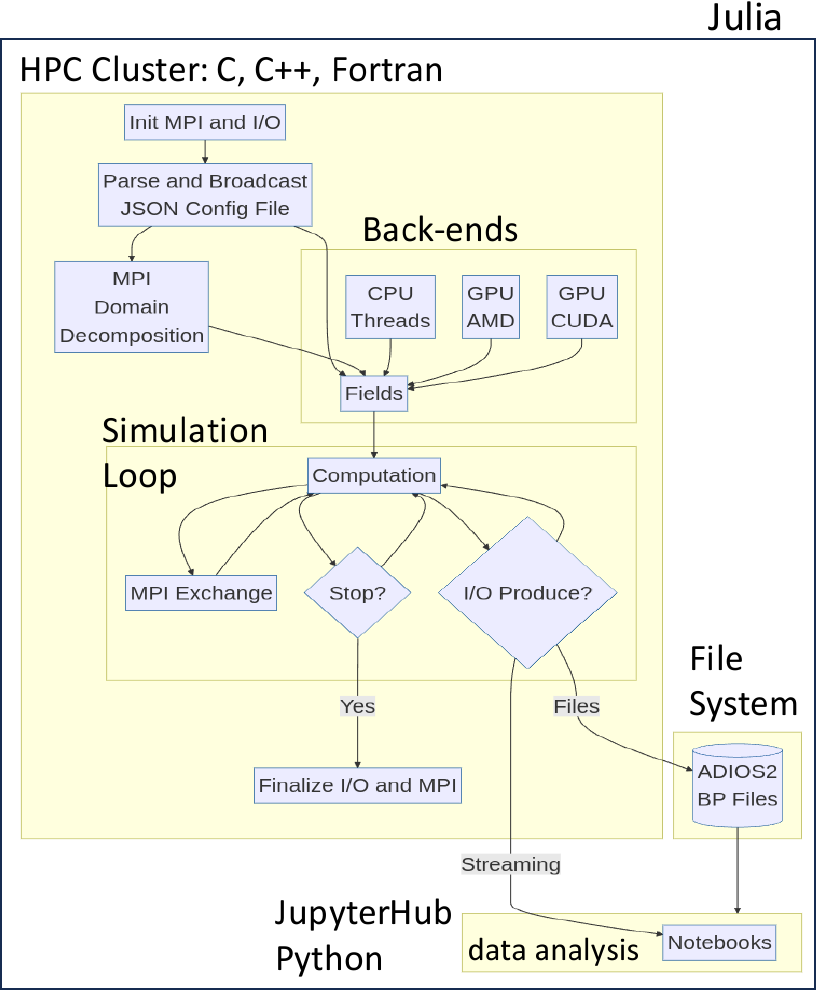}
  \caption{Gray-Scott workflow components schematic illustrating the unifying nature of Julia.}
  \label{fig:workflow}
\end{figure}

Figure~\ref{fig:workflow} shows a schematic representation of a typical parallel simulation and data analysis workflow components that apply to the current Gray-Scott use case. Julia's value proposition lies in its capacity to utilize a single language, along with an integrated ecosystem, to write these components. This unified approach allows the same language to not only operate at scale but also to conduct data analysis in an interactive environment, thereby fostering a seamless connection between computational execution and analytical exploration.

\subsection{Simulation}
\label{ssec:Simulation}

Gray-Scott is a two-variable, diffusion-reaction 3D model described by the PDEs shown in Equations (\ref{eqn:U}) and (\ref{eqn:V}):

\begin{subequations}
\begin{align}
\frac{\partial U}{\partial t} = D_U \nabla^2 U  - U V^2 + F \left( 1-U \right) + n r    
\label{eqn:U} \\
\frac{\partial V}{\partial t} = D_V \nabla^2 V  + U V^2 + - \left( F + k \right) V 
\label{eqn:V}
\end{align}
\end{subequations}

\noindent
where $U$ and $V$ are the output concentrations of two reacting and diffusing chemicals, while the inputs are listed as follows: 

\begin{itemize}
\item $D_u$ and $D_v$ are the diffusion rates for $U$ and $V$;
\item $F$ is the feed rate of $U$ into the system;
\item $k$ is the kill rate of $V$ from the system;
\item $n$ is the magnitude of the noise to be added to the system; and 
\item $r$ is a uniformly distributed random number between $-$1 and 1 for each time and spatial coordinate.
\end{itemize}

As illustrated in Equations~(\ref{eqn:Udiscretized}) and (\ref{eqn:Vdiscretized}), the set of governing equations is discretized in time, $t$, and space, $i,j,k$, on a regular normalized mesh by using simple forward and central differences, respectively:

\begin{subequations}
\begin{align}
U^{t+1}_{i,j,k} &=& U^t_{i,j,k} + \Delta t \left [ D_U \nabla^2 U^t_{i,j,k} + S^t_U\right ]
\label{eqn:Udiscretized} \\
V^{t+1}_{i,j,k} &=& V^t_{i,j,k} + \Delta t \left [ D_V  \nabla^2 V^t_{i,j,k} + S^t_V \right ]
\label{eqn:Vdiscretized}
\end{align}
\end{subequations}

\noindent
where $\Delta t$ is an input time step variable, $S$ represents the local source terms for $U$ and $V$ defined in Equations~(\ref{eqn:U}) and (\ref{eqn:V}), and the Laplacian operator, $\nabla^2$, is defined in Equation~(\ref{eqn:laplacian}) for the 3D nearest-neighbor Jacobi 7-point stencil in normalized spatial units:

\begin{align}
\nabla^2 U^t_{i,j,k} = - \,U^t_{i,j,k} + \frac{1}{6} 
\left [ U^t_{i-1,j,k} + U^t_{i+1,j,k} + \right . \nonumber \\ 
        U^t_{i,j-1,k} + U^t_{i,j+1,k} + \nonumber \\  \left . U^t_{i,j,k-1} + U^t_{i,j,k+1} \right ] 
\label{eqn:laplacian}
\end{align}

\begin{figure}[!ht]
  \centering
  \includegraphics[width=0.75\columnwidth,height=0.65\columnwidth]{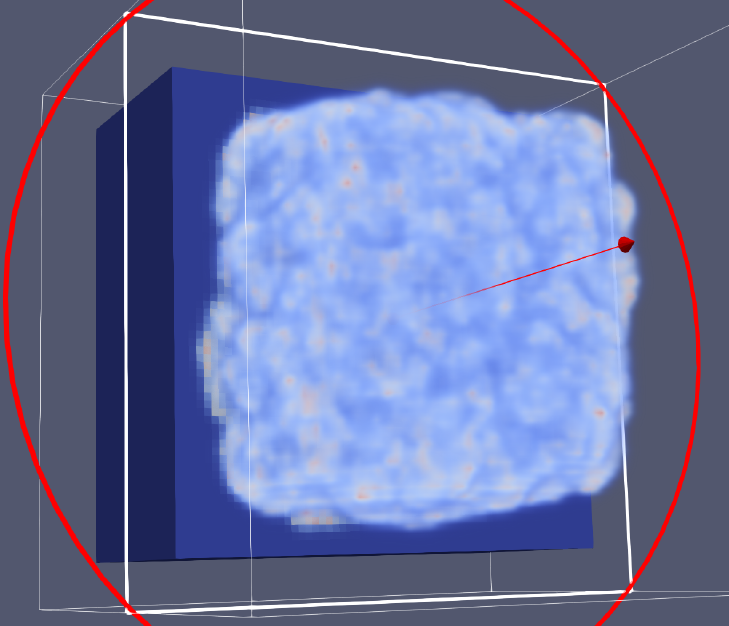}
  \caption{Gray-Scott snapshot of the solution of the 3D field $V$ variable showing the random nature of the source terms in the PDE.}
  \label{fig:GrayScottU}
\end{figure}

The random nature of the source term caused by $r$ and the coupled nature of $U$ and $V$ provide slightly more complexity than a typical diffusion calculation (e.g., heat transfer equation), as shown in Figure~\ref{fig:GrayScottU} for the visualization of the solution of the field variable $U$ at a center plane over a few time iterations. 

\subsection{On-Node Parallelization}

The stencil calculation is the main computational kernel in Gray-Scott. Stencil calculations are an important class of PDE solvers on structured grids~\cite{5222004}, and they have been successfully optimized for different hardware architectures, including GPUs \cite{10.1145/2304576.2304619,KROTKIEWSKI2013533,doi:10.1080/10407790.2010.541359}.

\begin{figure*}[!ht]
\centering
\includegraphics[width=0.9\linewidth]{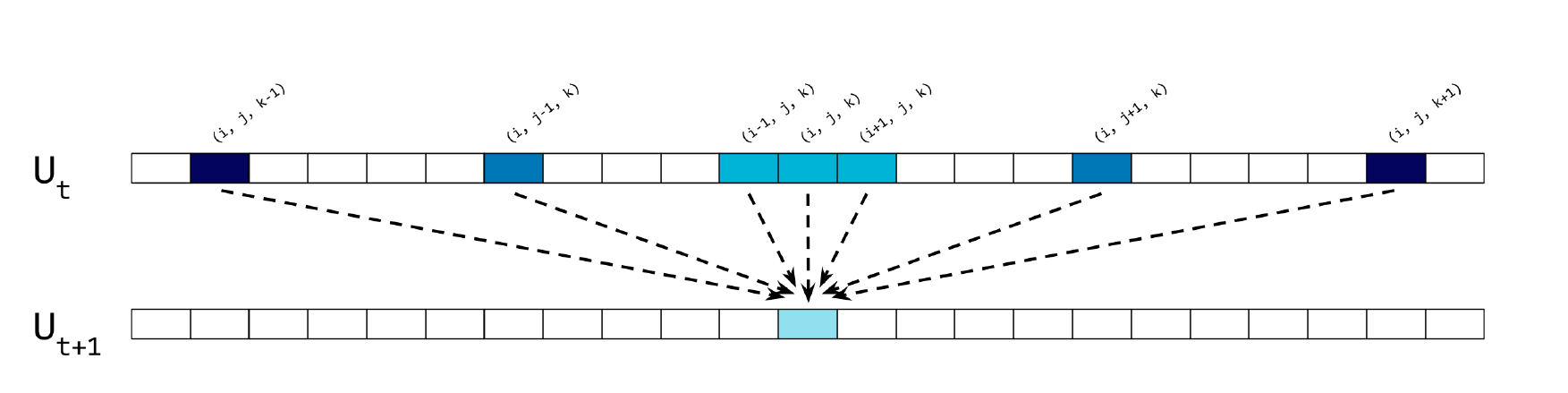}
\caption{Seven-point stencil memory access for one variable in the parallel Gray-Scott solver.}
\label{fig:memory}
\end{figure*}

Because each new time step value for $U$ and $V$ only depends on neighboring values, the problem is parallelizable on GPUs---and CPUs---thereby providing an extra memory allocation for temporary $U_{t+1}$ and $V_{t+1}$ field values for the next time iteration. Figure~\ref{fig:memory} illustrates the underlying contiguous memory allocated for each variable, which is accessed from near and far away points in memory; hence, some cache misses are expected in this typical kernel. Each variable requires 7 reads (fetch) $+$ 1 write $=$ 8 floating point access operations per variable for a total of 16 per cell in solving Equations (\ref{eqn:Udiscretized}) and (\ref{eqn:Vdiscretized}). 

\subsection{MPI Communication}

\begin{figure}[!ht]
\centering
\includegraphics[width=0.7\columnwidth,height=0.45\columnwidth]{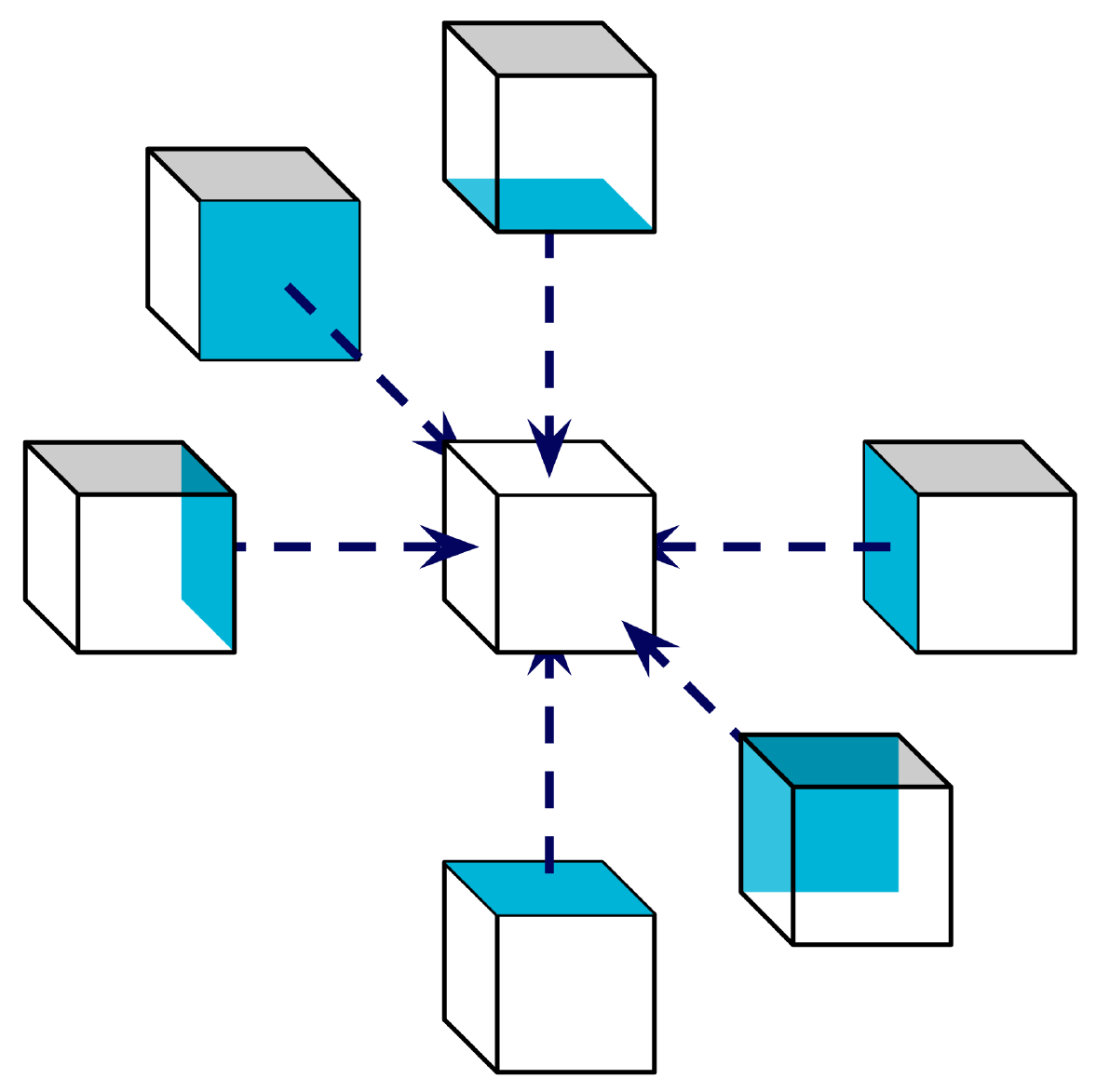}
\caption{MPI send/receive memory patterns in a Cartesian communicator decomposition used in Gray-Scott for each variable $U$ and $V$.}
\label{fig:CartComm}
\end{figure}

The 3D regular domain shown in Figure~\ref{fig:GrayScottU} is decomposed using an MPI Cartesian communicator. Each subdomain source shares face cell nodes with their neighboring subdomain destination at every computational iteration. This is illustrated in Figure~\ref{fig:CartComm}, which shows a typical \texttt{MPI\_Send}/\texttt{MPI\_Recv} pattern with \textit{ghost cell} surfaces. Because the exchanged surfaces are not memory contiguous, we define a new strided vector type by using \texttt{MPI\_Datatypes} and \texttt{MPI\_Type\_vector}~\cite{wu2016gpu}. We keep the MPI exchange from CPU-allocated memory to populate the strided vector contents coming from the GPU. We did not experiment with GPU-aware MPI for this work.

\subsection{Parallel I/O and Data Analysis}

Data production and consumption is performed to enable post-processing (e.g., visualization) of the solution of the $U$ and $V$ field variables over time. As such, a simple visualization schema and the corresponding metadata accounting for the input provenance are stored in the resulting dataset. Due to the volume of data produced from the simulation, drastically reducing the frequency of writes to the parallel file system is often required. Listing~\ref{lst:data} presents a sample provenance record of the data produced from the Gray-Scott simulation.

\begin{lstlisting}[language=Julia,caption={Provenance of the data generated from the Gray-Scott simulation.},label={lst:data}]]
  double   Du       attr   = 0.2
  double   Dv       attr   = 0.1
  double   F        attr   = 0.02
  double   U        1000*{1024, 1024, 1024} 
                    Min/Max -0.120795 / 1.46671
  double   V        1000*{1024, 1024, 1024} 
                    Min/Max  0 / 0.959875
  double   dt       attr   = 1
  double   k        attr   = 0.048
  double   noise    attr   = 0.1
  int32_t  step     50*scalar = 20 / 1000
  Attribute visualization schemas: FIDES, VTX
\end{lstlisting}

\section{Implementation}
\label{sec:Implementation}

We developed GrayScott.jl as a comprehensive HPC workflow application packaged as a Julia module.
The resulting Julia implementation using the AMDGPU.jl \cite{AMDGPU} package is shown in Listing~\ref{lst:GrayScott}. This implementation is a straightforward application of a typical GPU kernel and maps each component of the kernel onto a 3D computational grid. When launching this kernel and allocating the number of threads and grid size, one must consider that Julia arrays are column-major. Therefore, the fastest index, being the first one, should be structured to avoid splitting across threads on the GPU. 

Although the goal is not to increase the efficiency of the present stencil calculation, we aim to evaluate how the performance of the JIT generated Julia LLVM-IR stencil code compares with theoretical bandwidths and a simple AMD-provided Laplacian kernel. The metrics for this implementation are described and provided in Section~\ref{sec:Results}.

\begin{lstlisting}[language=Julia,caption={Julia AMDGPU.jl Gray-Scott kernel.},label={lst:GrayScott}]]
using AMDGPU
using Distributions

function _laplacian(i, j, k, var)
  l =   var[i - 1, j, k] + var[i + 1, j, k]  
      + var[i, j - 1, k] + var[i, j + 1, k] 
      + var[i, j, k - 1] + var[i, j, k + 1] 
      - 6.0 * var[i, j, k]
  return l / 6.0
end
  
function _kernel_amdgpu!(u, v, u_temp, v_temp, 
                         sizes, Du, Dv, F, K, 
                         noise, dt)  
  
  k = (workgroupIdx().x - 1) * workgroupDim().x 
      + workitemIdx().x
  j = (workgroupIdx().y - 1) * workgroupDim().y 
      + workitemIdx().y
  i = (workgroupIdx().z - 1) * workgroupDim().z 
      + workitemIdx().z
        
  if k == 1 || k >= sizes[3] || 
     j == 1 || j >= sizes[2] || 
     i == 1 || i >= sizes[1]
    return
  end

  @inbounds begin
    u_ijk = u[i, j, k]
    v_ijk = v[i, j, k]

    du = Du * _laplacian(i, j, k, u) 
         - u_ijk * v_ijk^2 + F * (1.0 - u_ijk) 
         + noise * rand(Uniform(-1, 1))
               
    dv = Dv * _laplacian(i, j, k, v) 
         + u_ijk * v_ijk^2 - (F + K) * v_ijk

    u_temp[i, j, k] = u_ijk + du * dt
    v_temp[i, j, k] = v_ijk + dv * dt
  end

  return nothing
end
\end{lstlisting}

\vskip 1cm

For managing communication, we use the MPI.jl package, which provides a high-level binding interface with the underlying MPI back end. The vector data types and MPI send/receive communication code for a single variable ghost cell surface are described in Listing~\ref{lst:MPI}. MPI.jl uniquely enables referencing the internal memory through \texttt{@view} to form a buffer structure prior to an actual send/receive operation. We purposely introduce variations in the amount of data shared across the $x$,$y$, and $z$ dimensions to capture distinct communication patterns. 

\begin{lstlisting}[language=Julia,caption={GrayScott.jl MPI.jl implementation},label={lst:MPI}]]
# if in GPU memory move first to CPU
u = typeof(fields.u) <: Array ? fields.u : 
                                Array(fields.u)
                                
xy_face_t = MPI.Types.create_vector(size_y + 2, 
            size_x, size_x + 2, MPI.Datatype(T))
MPI.Types.commit!(xy_face_t)
...
send = MPI.Buffer(@view(u[2, 1, size_z + 1]), 
       1, data_type)
recv = MPI.Buffer(@view(u[2, 1, 1]), 
       1, data_type)
MPI.Sendrecv!(send, recv, comm, dest = rank1,
              source = rank2)
\end{lstlisting}

We use the ADIOS2 library via the Julia ADIOS2.jl bindings. 
ADIOS2 is an MPI-based library, and the variables are described by accounting for their partitioning in a parallel I/O in which each MPI process produces a \texttt{block} of data in the global $U$ and $V$ 3D array variables. A typical ADIOS2 dataset is shown in Listing~\ref{lst:data} in which the combination of all data producer blocks from the MPI subdomain form the global $U$ and $V$ arrays from the PDE solution. We add schema attributes to enable later visualization in ParaView~\cite{ayachit2015paraview} by using FIDES~\cite{pugmire2021fides} and VTX readers implemented in VTK~\cite{schroeder2000visualizing}.

\subsection{The Frontier Supercomputer}

Frontier is equipped with AMD CPUs and GPUs and achieved 1.194 EFLOPS in the HPL benchmark~\cite{dongarra2003linpack} according to the June 2023 TOP500 list.\footnote{\url{https://www.top500.org/lists/top500/2023/06/}} Table~\ref{tab:Frontier} lists Frontier system characteristics and the software stack used for this study. In structuring our numerical experiments, we have designed them to grow by a factor of 8, so each dimension will double ($2^3=8$), thereby aligning with the GPU's maximum limits of 1,024 threads per dimension in a 3D computational kernel placement (Listing~\ref{lst:GrayScott}).

\begin{table}[ht]
\caption{Summary of Frontier hardware and software used in this study.}
\small{
\begin{center}
\begin{tabular}{ |c|c| } 

 \hline
  \multicolumn{2}{|c|}{\textbf{Frontier Characteristics}}\\
 \hline
 Nodes & 9,408 \\
 \hline 
 \hline
 \multicolumn{2}{|c|}{---CPU Architecture---}\\
 \hline
 \hline
 \hline
 CPU & AMD EPYC 7A53\\
 Cores & 64\\
 Memory &  DDR4 512~GB\\
 Bandwidth & 205~GB/s\\
 \hline
 \hline
 \hline
 \multicolumn{2}{|c|}{---GPU Architecture---}\\
 \hline
 \hline
 \hline
 GPU & 4$\times$AMD MI250X\\
     & 8$\times$GCDs\\
 Memory & HBM2E 64~GB\\
 Bandwidth & 1,600 GB/s per GCD \\
 \hline
 \hline
 \hline
 \multicolumn{2}{|c|}{---Connectivity---}\\
 \hline
 \hline
 \hline
 GPU-to-GPU & Infinity Fabric\\
 Bandwidth & 50--100 GB/s\\
 %Bandwidth & 50 GB/s & 32 GB/s\\
 GPU-to-CPU & Infinity Fabric\\
 Bandwidth & 36~GB/s\\
 \hline
 \hline
 \hline
\multicolumn{2}{|c|}{---File System---}\\
 \hline
 \hline
 \hline
 Type &   Lustre Orion \\
 Capacity & 679 PB \\
 Nodes   & 40 metadata \\
         & 450 object storage service (OSS)\\
 Write speed &  5.5 TBps  \\
 Read speed  &  4.5 TBps  \\
 \hline
 \hline
 \hline
 \multicolumn{2}{|c|}{---Software versions---}\\
 \hline
 \hline
 \hline
 Julia  &   1.9.2 \\
 AMDGPU.jl &   0.4.15 \\
 ROCm     & 5.4.0 \\
 MPI.jl   & 0.20.12 \\
 Cray-MPICH & 8.1.23 \\
 ADIOS2.jl &   1.2.1 \\
 ADIOS2    &  2.8.3 \\
\hline
\hline
\end{tabular}
\end{center}
}
\vspace{-0.1in}
\label{tab:Frontier}
\end{table}
\section{Results and Discussion}
\label{sec:Results}

In this section, we present the results from our evaluation of the Gray-Scott HPC workflow application regarding (i) compute capabilities on Frontier's MI250x AMD GPU using AMDGPU.jl, (ii) weak scalability of the simulation wall-clock times using Julia's MPI.jl, and (iii) simulation data production and consumption rates in JupyterHub using Julia's bindings to ADIOS2, ADIOS2.jl.\footnote{\url{https://github.com/eschnett/ADIOS2.jl}}

\begin{figure*}[!ht]
  \centering \includegraphics[width=2\columnwidth,height=0.44\columnwidth]{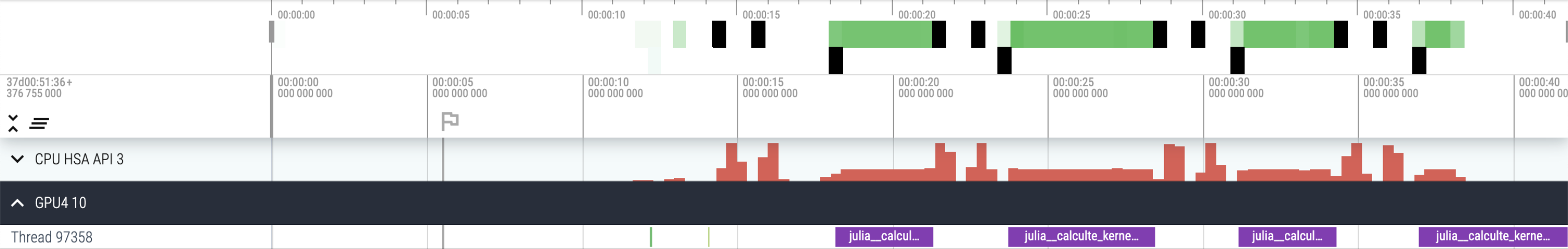}
  \caption{Gray-Scott simulation trace obtained with \texttt{rocprof} showing computational load on GPU and memory transfer to CPU for communication.}
  \label{fig:trace}
\end{figure*}

\subsection{Computation on MI250x GPUs}

As explained in Section~\ref{ssec:Simulation}, the metrics for the memory-bound Gray-Scott kernel presented in Listing~\ref{lst:GrayScott} are based on the performance of accessing the GPU allocated arrays $u$ and $v$ for 7 read (fetch/load) operations and 1 write operation. 
We collect two types of bandwidth information: (i)~effective bandwidth, meaning the expected minimal data movement (fetch/load and write/store) according to the number of operations in a single domain, and (ii) the total measured bandwidth obtained from AMD's \texttt{rocprof} profiler for \texttt{FETCH\_SIZE} and \texttt{WRITE\_SIZE}, which represent the total number of KB moved to and from the GPU memory (Figure~\ref{fig:trace}).

The effective fetch and write sizes can be estimated from the number of floating point operations per cell by considering that edge cells execute fewer operations due to reduced stencil size. A regular grid with $L$ number of cells per direction is provided in Equations (\ref{eqn:fetch}) and (\ref{eqn:write}), 
as reported in AMD's training material.\footnote{\url{https://github.com/AMD/amd-lab-notes}}

\begin{subequations}
\begin{align}
fetch\_size_{effective} = \left [ L^3 - 8 - 12 \left (L - 2 \right ) \right ] \cdot sizeof(T) \label{eqn:fetch} \\
write\_size_{effective} = (L - 2)^3 \cdot sizeof(T) \label{eqn:write}
\end{align}
\end{subequations}

Hence, the effective and total bandwidth metrics are shown in 
Equations (\ref{eqn:beff}) and (\ref{eqn:btot}) for the effective (theoretical) and the actual (measured with \texttt{rocprof}) GPU data access bandwidths.

\begin{subequations}
\begin{align}
bandwidth_{effective} = \frac{ \left( fetch\_size + write\_size \right)_{effective} }{kernel\_time} 
\label{eqn:beff} \\
bandwidth_{total} = \frac{ \left( FETCH\_SIZE + WRITE\_SIZE \right)_{rocprof} }{kernel\_time} 
\label{eqn:btot}
\end{align}
\end{subequations}

\begin{table}[!ht]
\centering
\caption{Average bandwidth comparison of different stencil implementations on a single GPU.}
\begin{tabular}{|l|r|r|}
 \hline
 Kernel &  \multicolumn{2}{c|}{Bandwidth (GB/s) }  \\
        &  Effective & Total \\
 \hline 
 Julia GrayScott.jl   &   &\\
 - 2-variable (application)  &  312  & 570 \\
 - 1-variable no random      & 312 & 625 \\
 \hline
 HIP single variable &  599  & 1,163 \\
 Theoretical peak MI250x        & \multicolumn{2}{c|}{1,600}\\
 \hline \hline
\end{tabular}
\label{tab:MI250x}
\end{table}

\begin{table}[ht]
\centering
\caption{The \texttt{rocprof} outputs for HIP 1-variable and Julia Gray-Scott.jl implementations.}
\begin{tabular}{|l|r|r|r|}
\hline
Kernel &  HIP   & \multicolumn{2}{c|}{GrayScott.jl}  \\ 
metric &  1-var & 1-variable    & 2-variable \\
       &        & no random     & (application) \\
   \hline
wgr & 256 & 512 & 512 \\
lds & 0   & 29,184 & 29,184 \\
scr & 0   &  8,192 &  8,192 \\
FETCH\_SIZE (GB) & 25.08 & 25.40 & 50.80 \\
WRITE\_SIZE (GB) & 8.35  & 8.38  & 16.78 \\
TCC\_HIT (M) &  9.14   & 10.80    & 24.60 \\  
TCC\_MISS (M) &  8.36  & 8.69 &  17.19 \\
Avg Duration (ms) & 28.74  & 54.03 & 111.07 \\
\hline \hline
\end{tabular}
\label{tab:rocprof}
\end{table}

Listing~\ref{lst:LLVM-IR} presents the generated LLVM-IR memory accesses in the Julia implementation. These numbers are consistent with the algorithm patterns in Listing~\ref{lst:GrayScott} of 16 loads and 2 stores. Consequently, we did not observe additional overhead typical of high-level implementations. The reasons for the bandwidth performance difference in Table~\ref{tab:rocprof} are beyond the IR level and might present opportunities for lower-level optimizations at the LLVM instruction set architecture level, thereby providing an opportunity to close gaps from vendor-specific optimizations as Julia's AMDGPU.jl and the AMD HIP/ROCm stack continues to rapidly evolve.

\begin{lstlisting}[language=LLVM,caption={GrayScott.jl application kernel unique memory loads (14) and store (2) in LLVM-IR.},label={lst:LLVM-IR},basicstyle=\ttfamily\tiny]
       %94 = load double, double addrspace(1)*  %93, align 8
      %103 = load double, double addrspace(1)* %102, align 8
      %107 = load double, double addrspace(1)* %106, align 8
      %110 = load double, double addrspace(1)* %109, align 8
      %114 = load double, double addrspace(1)* %113, align 8
      %117 = load double, double addrspace(1)* %116, align 8
      %122 = load double, double addrspace(1)* %121, align 8
      %126 = load double, double addrspace(1)* %125, align 8
      %312 = load double, double addrspace(1)* %311, align 8
      %315 = load double, double addrspace(1)* %314, align 8
      %318 = load double, double addrspace(1)* %317, align 8
      %321 = load double, double addrspace(1)* %320, align 8
      %325 = load double, double addrspace(1)* %324, align 8
      %329 = load double, double addrspace(1)* %328, align 8
      ...
      store double %345, double addrspace(1)* %353, align 8
      store double %355, double addrspace(1)* %363, align 8
\end{lstlisting}

\subsection{Weak scaling}

\begin{figure}[!ht]
\centering
\includegraphics[width=1.\columnwidth,height=0.7\columnwidth]{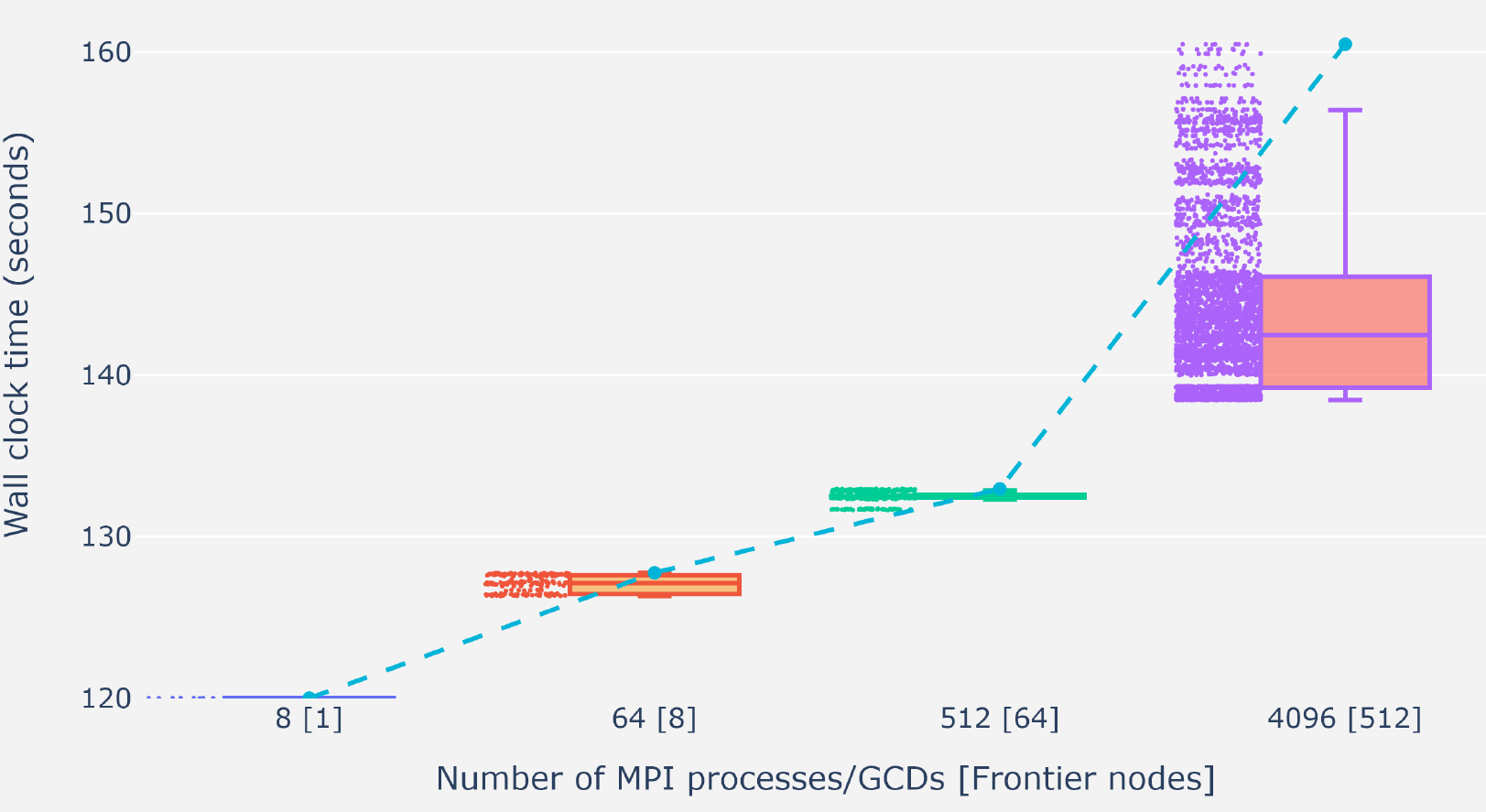}
\caption{Weak scaling with single MPI process wall-clock time variability obtained with Gray-Scott.jl on Frontier.}
\label{fig:weak}
\end{figure}

We ran GrayScott.jl with different cell sizes using 1 GCD (GPU) per MPI process. Weak scaling results are illustrated in Figure~\ref{fig:weak} and show the total wall-clock times per MPI process for a constant workload of $nx = ny = nz = 1,024$ per dimension, or $1,024^3 = 1,073\times10^9$ cells per GPU. 

Per-process wall-clock time variability in HPC workflows is proportional to the number of MPI processes. As expected, the overall communication overhead is dictated by the slowest time-to-solution processes (dashed blue line). The Julia code could run successfully on 4,096 GPUs or 512 nodes of Frontier (5.44\% of the system). A clear trend of small variability of a few percent (2\%--3\%) is apparent up to 512 MPI processes, whereas large variability (12\%--15\%) appears for the largest case. Although we do not have a baseline comparison, this initial measurement can help with further optimizations because Julia can produce realistic scenarios for MPI communication on a system such as Frontier. When attempting runs for the next size for a factor 8 at 32,768 GPUs (4,096 nodes, 50\% of the Frontier system), unpredictable failures occurred at the underlying MPI layers during the ghost cell exchange stage. Nevertheless, all 32,768 GPUs showed initial runs keeping the bandwidth close to the expected value of 312 GB/s, as reported in Table~\ref{tab:MI250x}.

\begin{figure}[!ht]
  \centering \includegraphics[width=1\columnwidth,height=0.6\columnwidth]{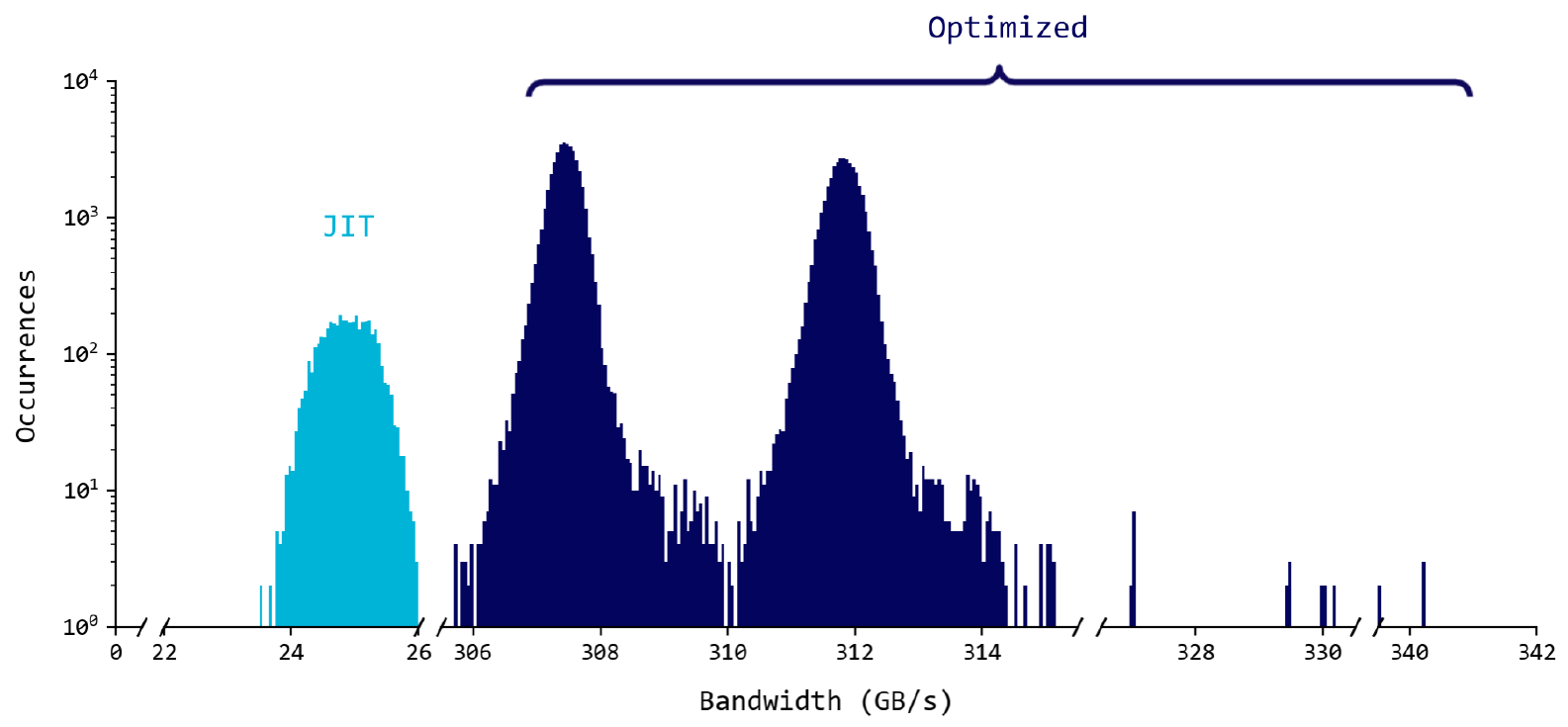}
  \caption{Bandwidth distribution for JIT and optimized kernel code running on 4,096 Frontier GPUs for 20 simulation steps.}
  \label{fig:bandwidths}
\end{figure}

In the context of HPC workflows, understanding the interplay between scale and performance is critical. To this end, we delve deeper into GPU bandwidth metrics from large-scale runs. Figure~\ref{fig:bandwidths} shows the obtained bandwidth distributions for the initial JIT compiled run, which represents the initial overhead in running Julia kernels, and the optimized code when running on 4,096 GPUs. The initial JIT compilation represents the early overhead associated with running Julia kernels in an HPC workflow---an $\approx$12.5$\times$ cost, or 8\% on average, of the bandwidth in the optimized kernel. Although JIT compilation represents an amortized cost, it should be incorporated in the estimation of total wall-clock run times. While this overhead can be addressed in HPC workflow planning, Julia's ahead-of-time mechanism was not explored in this study. These insights not only demonstrate Julia's capacity for HPC workflows but also reveal areas for further research and optimization, which could enable more efficient utilization of resources in complex computational environments.

\subsection{Parallel I/O}

\begin{figure}[!ht]
\centering
\includegraphics[width=1\columnwidth,height=0.7\columnwidth]{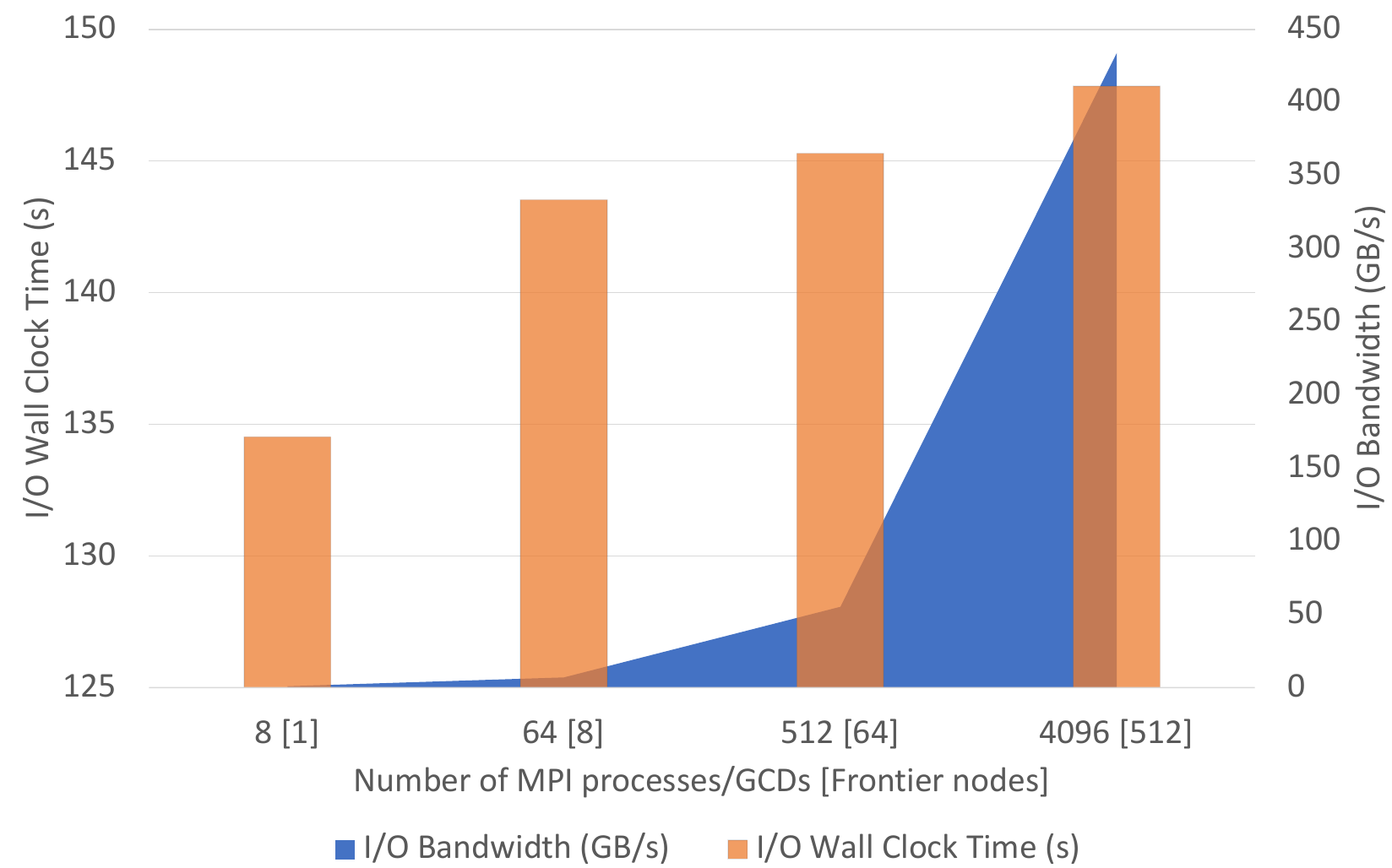}
\caption{Weak scaling on parallel I/O showing wall-clock times and bandwidth performance when using ADIOS2.jl on Frontier.}
\label{fig:parallel-io}
\end{figure}

In HPC workflows, the efficient use of parallel file systems is crucial for both saving computational results and ensuring smooth data transfer between different stages of a workflow. To this end, we evaluate using the parallel file system within the GrayScott.jl package, focusing on saving one output step of the cases presented in Figure~\ref{fig:weak}. Results from using the ADIOS2.jl Julia bindings package are presented in Figure~\ref{fig:parallel-io}, revealing the performance characteristics of weak scaling in the writing of the dataset described in Section~\ref{sec:Implementation}. The underlying ADIOS2 library uses the default BP5 format, which generates a single sub-file per node. At 512 nodes, write times are fairly flat, which represents an increase of write bandwidth up to 434~GB/s for the largest case. This represents an 8\% peak file system write performance of 5.5~TB/s when using 5\% of the machine (Table~\ref{tab:Frontier}). In the context of HPC workflows, such a performance metric is significant, as parallel I/O performance is dictated by low-level POSIX write operations, the ADIOS2 buffering management, and real-time file system usage. With the Julia binding, we observed negligible overhead, indicating that Julia can serve as an effective tool in HPC workflow optimization. 

\begin{figure}[!ht]
\centering
\includegraphics[width=\columnwidth,height=1.6\columnwidth]{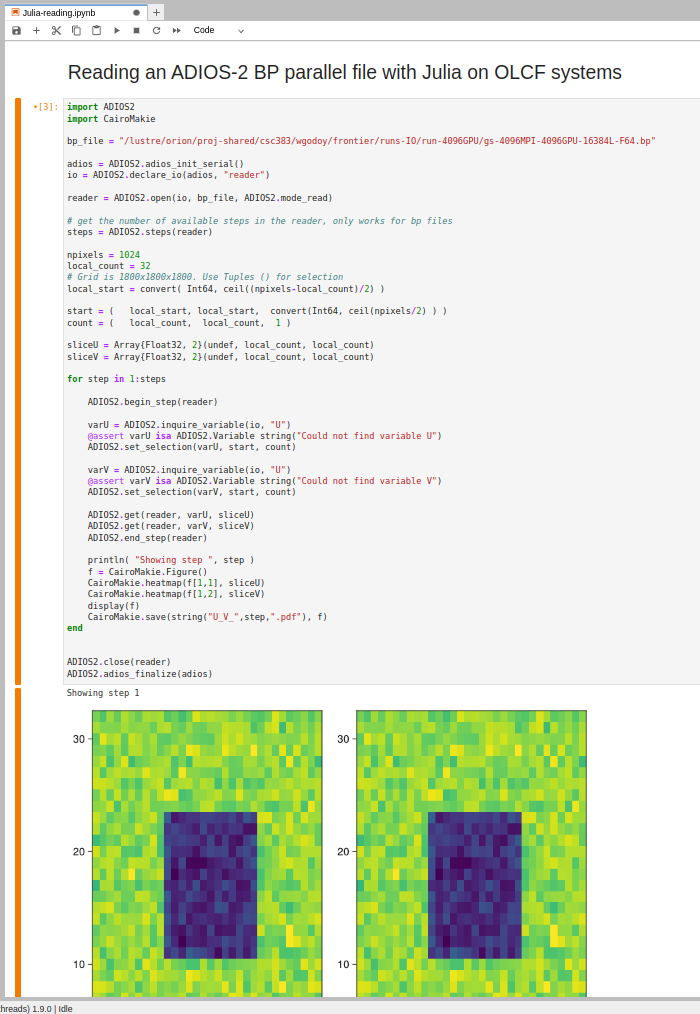}
\caption{Snapshot of Julia data analysis implementation on the OLCF's JupyterHub server.}
\label{fig:jupyterhub}
\end{figure}

Finally, we look at Julia in an end-to-end workflow development process for data analysis. We tested a Julia implementation by 
reading a plotting of simple 2D slices of the stored $U$ and $V$ 3D arrays with the Makie.jl library~\cite{DanischKrumbiegel2021}. Figure~\ref{fig:jupyterhub} shows a snapshot of the resulting JupyterHub service consuming the ADIOS2-generated data shared on Frontier's file system. This instance demonstrates Julia's potential as a unified language that can seamlessly connect different stages of an HPC workflow, from computational simulation to data visualization. Future work includes further performance tuning and evaluation of the trade-offs for in-memory streaming data pipelines~\cite{10.1007/978-3-030-96498-6_6}.

By bridging various components in a typical HPC workflow, Julia could emerge as a compelling option for scientists and engineers working on complex, large-scale computational problems. It also sets the stage for more integrated, responsive, and adaptive HPC workflows that can fully leverage the capabilities of modern supercomputing infrastructures.

\section{Conclusions}
\label{sec:Conclusions}

On up to 4,096 GPUs and MPI processes (512 nodes of the Frontier supercomputer), we used a 2-variable diffusion-reaction code, Gray-Scott, to test the performance of the Julia HPC ecosystem in the development of tightly coupled workflow components. Overall, the Julia stencil solver achieves close to a 50\% bandwidth on Frontier's MI250x AMD GPUs vs. a HIP implementation; hence, performance gaps still exist and must be closed as we look forward to future versions of the actively developed AMDGPU.jl. Meanwhile, the measured weak scaling from the MPI communication and parallel I/O components suggest that bindings available in Julia are lightweight and perfectly suitable running on top of the underlying MPI and ADIOS2 library implementations. The Julia implementation shows similar patterns in overhead and variability typical of network and file system communication in HPC systems. We also showcased the data analysis feasibility on available JupyterHub systems.
Overall, the LLVM-based Julia HPC ecosystem presents an attractive alternative for developing co-design components given the high-performance and high-productivity requirements for the end-to-end workflows that power scientific discovery (e.g., AI, FAIR).

\begin{acks}
This research was supported by the Exascale Computing Project (17-SC-20-SC), a collaborative effort of the US Department of Energy Office of Science and the National Nuclear Security Administration. This research used resources of the Oak Ridge Leadership Computing Facility and the Experimental Computing Laboratory (ExCL) at the Oak Ridge National Laboratory, which is supported by the Office of Science of the US Department of Energy under Contract No. DE-AC05-00OR22725. This work is funded, in part,  by Bluestone, a X-Stack project in the DOE Advanced Scientific Computing Office with program manager Hal Finkel.
\end{acks}

\bibliographystyle{ACM-Reference-Format}
\bibliography{references}

\appendix

\section{Artifact Description}
\label{ap1:Artifact}

The GrayScott.jl implementation is available on GitHub: 
\url{https://github.com/JuliaORNL/GrayScott.jl}.

The repository contains instructions for the code and how to run it locally, including setting up the JSON configuration files used in this study, which are also available: \\
\url{https://github.com/JuliaORNL/GrayScott.jl/blob/main/examples/settings-files.json}. 

To run the codes on Frontier, we created a specific branch to avoid issues related to CUDA.jl and random number generation on AMDGPU.jl until fixes are available: \\ \url{https://github.com/JuliaORNL/GrayScott.jl/tree/amdgpu-frontier}.

We provide a configuration file to set up the modules environment as of June/July 2023: \url{https://github.com/JuliaORNL/GrayScott.jl/blob/amdgpu-frontier/scripts/config_frontier.sh}.

We used the latest available release of Julia, version 1.9.2.

\end{document}